\documentclass[a4paper,draft]{agujournal2019}
\usepackage[T1]{fontenc}
\usepackage[latin9]{inputenc}
\usepackage{units}
\usepackage{url}
\usepackage{amsbsy}
\usepackage{amstext}
\usepackage{graphicx}
\usepackage{rotfloat}
\usepackage{esint}
\usepackage[authoryear]{natbib}

\makeatletter

\pdfpageheight\paperheight
\pdfpagewidth\paperwidth

\journalname{Journal of Geophysical Research: Space Physics}

\usepackage{soul}

\usepackage{caption}
\makeatother

\begin{document}

\title{How a realistic magnetosphere alters the polarizations of surface,
fast magnetosonic, and Alfv\'{e}n waves}

\authors{M. O. Archer, \affil{1}\\
D. J. Southwood, \affil{1}\\
M. D. Hartinger, \affil{2}\\
L. Rastaetter, \affil{3}\\
and A. N. Wright \affil{4}}

\affiliation{1}{Space and Atmospheric Physics Group, Department of Physics, Imperial
College London, London, UK.}

\affiliation{2}{Space Science Institute, Boulder, Colorado, USA.}

\affiliation{3}{NASA Goddard Space Flight Center, Greenbelt, Maryland, USA.}

\affiliation{4}{Department of Mathematics and Statistics, University of St Andrews,
St Andrews, UK.}

\correspondingauthor{Martin Archer}{m.archer10@imperial.ac.uk}

\begin{keypoints}

\item A global MHD simulation shows system-scale ULF waves' polarizations
can significantly differ from box and dipole model predictions

\item Phase or handedness reversals in the magnetic field compared
to the velocity can occur simply due to the highly non-uniform background
field

\item We propose modified detection techniques for spacecraft observations
which account for the effects of a realistic magnetosphere

\end{keypoints}
\begin{abstract}
System-scale magnetohydrodynamic (MHD) waves within Earth\textquoteright s
magnetosphere are often understood theoretically using box models.
While these have been highly instructive in understanding many fundamental
features of the various wave modes present, they neglect the complexities
of geospace such as the inhomogeneities and curvilinear geometries
present. Here we show global MHD simulations of resonant waves impulsively-excited
by a solar wind pressure pulse. Although many aspects of the surface,
fast magnetosonic (cavity/waveguide), and Alfv\'{e}n modes present
agree with the box and axially symmetric dipole models, we find some
predictions for large-scale waves are significantly altered in a realistic
magnetosphere. The radial ordering of fast mode turning points and
Alfv\'{e}n resonant locations may be reversed even with monotonic
wave speeds. Additional nodes along field lines that are not present
in the displacement/velocity occur in both the perpendicular and compressional
components of the magnetic field. Close to the magnetopause the perpendicular
oscillations of the magnetic field have the opposite handedness to
the velocity. Finally, widely-used detection techniques for standing
waves, both across and along the field, can fail to identify their
presence. We explain how all these features arise from the MHD equations
when accounting for a non-uniform background field and propose modified
methods which might be applied to spacecraft observations.
\end{abstract}

\section*{Plain Language Summary}

Earth's magnetic environment in space, the magnetosphere, is a complex
system within which ultra-low frequency analogues to sound waves in
the space plasmas present form various different types of resonance,
somewhat like in musical instruments. Our understanding of such oscillations
come from highly simplified mathematical models, which neglect many
aspects of reality as they are difficult to include. By using computer
simulations of the magnetosphere, however, we can compare the results
from more representative conditions with our predictions from the
easier theory. We find that while many features of the different waves
present do indeed agree with expectations, some of the predictions
become significantly altered by the use of a realistic magnetosphere.
We explain how these differences arise and propose new techniques
that take them into account, which might be used by those analysing
measurements of these waves from orbiting satellites.

\section{Introduction}

\begin{figure}
\begin{centering}
\noindent \makebox[\textwidth]{\includegraphics{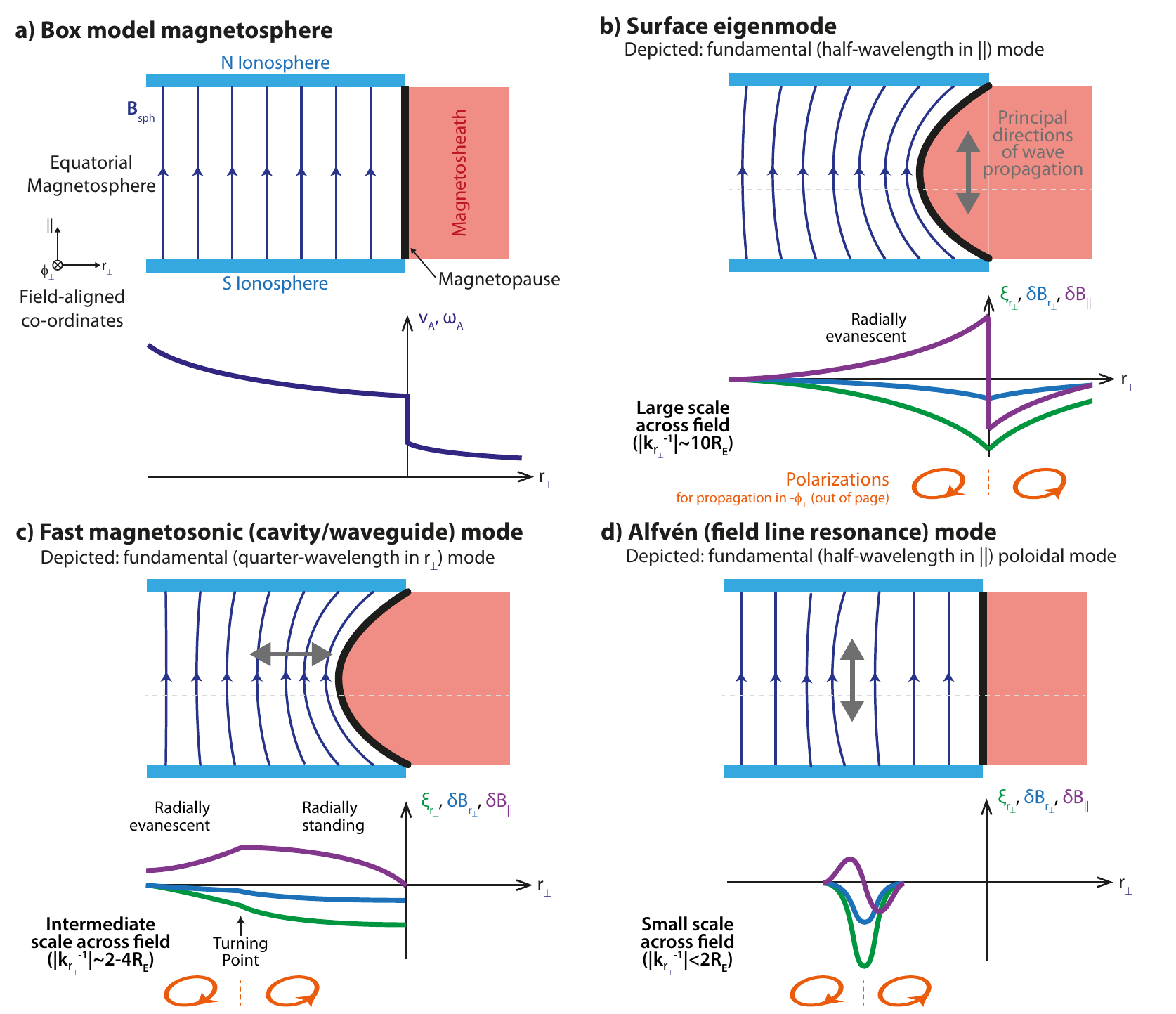}}
\par\end{centering}
\caption{Cartoon illustrating a) a box model magnetosphere and b--d) MHD wave
eigenmodes within it. Shown are a magnetopause surface eigenmode (b),
cavity/waveguide mode (c), and poloidal field line resonance (d).
Inset graphs show (a) a monotonic Alfv\'{e}n speed profile, (b--d)
instantaneous variations along the dotted lines of radial displacement
(green) and perpendicular (blue) and compressional (purple) magnetic
field perturbations. The sense of the polarisation is also shown in
orange, assuming westward (out of the page) propagating disturbances.\label{fig:modes-cartoon}}

\end{figure}

System-scale magnetohydrodynamic (MHD) waves in Earth's magnetosphere
are routinely observed by spacecraft and ground-based instrumentation
as ultra-low frequency (ULF) waves, with frequencies of fractions
of milliHertz to a few Hertz \citep{jacobs64}. These waves provide
a means for solar wind energy and momentum to be transferred throughout
geospace, e.g. to the radiation belts \citep{elkington06}. They can
also, through understanding the various normal modes they may establish,
be used as a tool for probing the ever-changing nature of the terrestrial
system \citep{menk13}. The foundations of MHD wave theory have largely
been built in so-called box models, where the curved geomagnetic field
lines are straightened into a uniform field anchored at the northern
and southern ionospheres due to their high conductivity as shown in
Figure~\ref{fig:modes-cartoon}a \citep[e.g.][]{radoski71,southwood74}.
While such analytic models have proven extremely useful, they are
of course highly simplified compared to reality, thus it is important
to understand their limitations. Numerical modelling of MHD waves
can help in this regard and there are a number of different approaches
which may be taken, from dedicated wave codes \citep[e.g.][]{lee99,degeling14,wright16}
to the use of general purpose global simulations \citep[e.g.][]{claudepierre10,hartinger14,ellington16}.
\citet{wright20} provide further discussion of the benefits/drawbacks
to each approach.

Comparing the polarization with theoretical predictions is an often
used method of deciphering the wave modes present within ULF wave
observations, both at Earth \citep[e.g.][]{samson71,mathie99,takahashi91,agapitov09,kokubun13}
and other planetary systems \citep[e.g.][]{james16,manners19}. Polarization
can refer to several different related aspects of a wave including:
the orientation of oscillations of a particular physical quantity,
the shape and handedness these trace out, relative amplitudes between
different quantities, or their cross-phases \citep{waters02}. When
waves have a definite sense of propagation azimuthally (assumed westwards
throughout Figure~\ref{fig:modes-cartoon}) their polarization depends
only on the gradient in amplitude across the field, which in typical
box and dipole models is either radially towards/away from the Earth,
and whether the waves are evanescent/propagating \citep{southwood74,southwood86}.
In the following subsections we briefly introduce the surface, fast
magnetosonic, and Alfv\'{e}n eigenmodes of the magnetosphere.

\subsection{Surface modes}

The dynamics of discontinuities in geospace may be described as surface
waves driven by upstream pressure variations or flow shears \citep{pu83,kivelson95}.
These have mostly been studied at the magnetopause flanks, where amplitudes
can grow via the Kelvin-Helmholtz instability \citep{fairfield00,otto05}.
Here the waves' frequency ($\omega/2\pi$) is largely controlled by
the magnetosheath velocity $\mathbf{v}_{msh}$, i.e. $\left|\omega-\mathbf{k}\cdot\mathbf{v}_{msh}\right|\approx k_{\phi}v_{msh}$
where $\mathbf{k}$ is the wavevector and $\phi$ the azimuthal direction
(see Notation). However, on the dayside, where flows are weaker, the
finite extent of magnetospheric field lines play a more significant
factor. \citet{chen74} proposed the possibility of surface eigenmodes
between conjugate ionospheres, only recently discovered observationally
at the magnetopause \citep{archer19} and plasmapause \citep{He2020}.
Figure~\ref{fig:modes-cartoon}b illustrates a magnetopause surface
eigenmode (MSE) in a box model \citep{plaschke11}, constructed as
evanescent fast magnetosonic waves either side of an infinitesimally
thin discontinuity. Surface waves therefore obey the usual fast wave
dispersion relation 
\begin{equation}
k_{r_{\perp}}^{2}=-k_{\phi_{\perp}}^{2}-k_{\parallel}^{2}+\frac{\omega^{4}}{\omega^{2}v_{A}^{2}+c_{s}^{2}\left(\omega^{2}-k_{\parallel}^{2}v_{A}^{2}\right)}\label{eq:magnetosonic-dispersion}
\end{equation}
where $v_{A}$ and $c_{s}$ are the Alfv\'{e}n and sound speeds respectively.
Incompressibility renders the last term of equation~\ref{eq:magnetosonic-dispersion}
negligible. The frequency is dictated by conditions either side

\begin{equation}
\omega_{MSE}=k_{\parallel}\sqrt{\frac{B_{sph}^{2}+B_{msh}^{2}}{\mu_{0}\left(\rho_{sph}+\rho_{msh}\right)}}\approx k_{\parallel}\frac{B_{sph}}{\sqrt{\mu_{0}\rho_{msh}}}\label{eq:MSE}
\end{equation}
for $k_{\phi_{\perp}}\ll k_{\parallel}$ \citep{plaschke09a}, which
predicts fundamental frequencies below $2\,\mathrm{mHz}$ \citep{archer15b}.
This makes MSE the lowest frequency magnetospheric normal mode and
highly penetrating. A finite thickness boundary is thought to damp
these collective modes through mode-conversion to oscillations within
the Alfv\'{e}n speed gradient, which undergo spatial phase-mixing
and dissipate energy to smaller (non-MHD) scales \citep{chen74,lee86,uberoi89}.
Whether this all occurs locally or if energy is deposited to the ionosphere
is not currently known.

While it is not understood theoretically how more realistic magnetic
geometries affect surface waves \citep{archer15,kozyreva19}, high-resolution
global MHD simulations have provided valuable insights. \citet{hartinger15},
henceforth H15, showed global $1.8\,\mathrm{mHz}$ waves excited by
a solar wind density pulse, consistent only with MSE. The amplitude
of the magnetic compressions/rarefactions decay with distance from
the magnetopause. However, inside the current layer two local maxima
occur, with a minimum between them near the peak current density.
The phase of the compressional magnetic field reverses either side
of the boundary, i.e. when the magnetosphere is compressed the magnetosheath
becomes rarefied. \citet{archer21}, A21 herein, found similar motion
of the subsolar bow shock, lagging behind the magnetopause. While
the magnetopause waves travel tailward at the equatorial flanks, between
$09h$ and $15h$ Magnetic Local Time (MLT) they are stationary despite
significant magnetosheath flows being present. The authors show the
time-averaged Poynting flux inside the magnetosphere surprisingly
points towards the subsolar point, perfectly balancing advection by
the magnetosheath flow such that there is no net (Poynting plus advective)
flux. Inside the magnetosphere, despite decaying amplitudes with distance,
phase fronts slowly propagate towards the magnetopause due to damping.
Finally, the Kelvin-Helmholtz instability causes seeded tailward propagating
surface waves to grow in amplitude.

\subsection{Fast magnetosonic (cavity/waveguide) modes}

Fast magnetosonic waves may form radially standing waves due to reflection
by boundaries (such as the magnetopause) or turning points (where
$k_{r_{\perp}}$ from equation~\ref{eq:magnetosonic-dispersion}
becomes zero) \citep{kivelson84,kivelson85}. These are known as cavity
modes in closed geometries \citep{allan86a} or waveguides when the
magnetosphere is open-ended \citep{samson92,wright94}. Azimuthal
wavenumbers are thus continuous in the latter but quantized in the
former. Many types of cavity/waveguide modes are known such as plasmaspheric,
virtual, tunnelling and trapped modes \citep{waters00}, with Figure~\ref{fig:modes-cartoon}c
depicting one between the magnetopause and a turning point, beyond
which the wave is evanescent.

The fast eigenmodes can be estimated under the Wentzel--Kramers--Brillouin
(WKB) approximation (full numerical solutions differ by only $\sim3\%$;
\citealp{rickard95}) by spatially integrating the radial wavenumber
(equation~\ref{eq:magnetosonic-dispersion}) and imposing a quantisation
condition \citep{samson92,samson95}. The fundamental mode was originally
thought to be a half-wavelength mode, with a node in displacement
at the magnetopause \citep{kivelson85,samson92}. \citet{mann99}
later showed that quarter-wavelength modes with a displacement antinode
at the boundary may be possible, which is what is shown in Figure~\ref{fig:modes-cartoon}c.
Both these modes have been successfully reproduced around noon within
global MHD simulations \citep{claudepierre09,hartinger14}. Cavity/waveguide
modes' structure and frequencies are thought to be highly dependent
on the Alfv\'{e}n speed profiles present \citep{allan89,wright95,archer15b,archer17},
though in general have higher frequencies and less penetrating scales
than surface modes. While numerical works suggest they should have
clear compressional magnetic field signatures with nodal structure
radially \citep{waters02,elsden19,elsden18}, identifying them in
satellite observations can be challenging \citep{hartinger12,hartinger13}.

\subsection{Alfv\'{e}n modes}

The final mode concerns Alfv\'{e}n waves standing along geomagnetic
field lines \citep{dungey67}. Often these occur over a range of $L$-shells
with a continuum of resonant frequencies present, however, sometimes
a discrete field line resonance is established \citep[e.g.][]{plaschke08},
as depicted in Figure~\ref{fig:modes-cartoon}d. Alfv\'{e}n modes
are typically described in terms of either poloidal or toroidal polarization.
In an axially symmetric dipole the toroidal mode corresponds to azimuthal
displacements of the plasma (and thus do not lead to magnetic compressions)
whereas poloidal modes feature radial ones (we note compressions become
negligible for high azimuthal wavenumbers though). WKB methods predict
no difference in frequencies between the two orientations, with the
fundamental given by

\begin{equation}
\omega_{A}\approx2\pi\left[2\int\frac{ds}{v_{A}}\right]^{-1}\label{eq:FLRfreq}
\end{equation}
\citet{singer91}, however, derived the wave equation within a general
orthogonal magnetic geometry

\begin{equation}
\frac{\partial^{2}}{\partial s^{2}}\left(\frac{\xi_{\alpha}}{h_{\alpha}}\right)+\frac{\partial}{\partial s}\left(\ln\left[h_{\alpha}^{2}B_{0}\right]\right)\frac{\partial}{\partial s}\left(\frac{\xi_{\alpha}}{h_{\alpha}}\right)+\frac{\omega^{2}}{v_{A}^{2}}\left(\frac{\xi_{\alpha}}{h_{\alpha}}\right)=0\label{eq:singer}
\end{equation}
where $\alpha$ represents some direction perpendicular to the background
field and $h_{\alpha}$ is its corresponding scale factor, estimated
as the distance between adjacent field lines. Numerical solutions
to this equation predict lower frequencies for the poloidal mode than
the toroidal one, which have been verified within simulations \citep[e.g.][]{elsden20}.
However, orthogonal coordinates only exist in the absence of background
field-aligned currents \citep{salat2000} and improvements which do
not require an orthogonal system have also been developed \citep{rankin06,kabin07,degeling2010}.
These have shown that the orientations of the two polarizations can
be altered by the local magnetic geometry. Asymmetries in the Alfv\'{e}n
speeds can also have a similar effect \citep{wright20}. Field line
resonances can also be reproduced in global MHD simulations \citep{claudepierre10,ellington16}.
The width of a discrete field line resonance is given by the length
scale of radial changes in the eigenfrequency \citep{southwood87,mann95},
which typically gives much shorter scales than the other two modes.

\subsection{Preface}

The three wave modes do not exist in isolation. In box and axially
symmetric dipole model setups wave coupling depends on the azimuthal
wavenumber, with no coupling predicted in the limits of zero or infinity
\citep{kivelson85,chen89}. However, in more realistic geometries
coupling is always expected \citep{radoski71}. Typically this is
discussed as the surface \citep{southwood74} or cavity/waveguide
\citep{kivelson84,kivelson85} mode exciting a field line resonance
at the radial location where their eigenfrequencies match. This theory,
however, is usually one-dimensional in nature and the problem of wave
coupling in a 3D asymmetric magnetosphere remains a topic of current
research. While dedicated MHD wave codes have shown progress in the
area of fast--Alfv\'{e}n mode coupling, only global MHD simulations
can self-consistently incorporate magnetopause surface modes too.
Therefore, in this paper we study one such simulation run to determine
how wave polarizations may be altered in a realistic magnetosphere
compared to the simplified box models.

\section{Simulation}

This paper uses a high-resolution Space Weather Modeling Framework
(SWMF; \citealp{toth05,toth12}) of the magnetospheric response to
a $1\,\mathrm{min}$ solar wind density pulse with sunward normal,
pressure-balanced with the ambient plasma via reduced temperature.
The interplanetary magnetic field (IMF) is kept constant and northward,
since this is most conducive to surface eigenmodes and the Kelvin-Helmholtz
instability \citep{southwood68,hasegawa75,plaschke11}. No plasmasphere
or ring current are included. The ionospheric conductivity is uniform
and the dipole is fixed with zero tilt throughout. The simulation
hence is both North--South and dawn--dusk symmetric. Full details
of parameters used are found in Table~S1. This specific simulation
run was first presented by A21 which in turn was essentially a replication
on NASA's Community Coordinated Modeling Center (CCMC) of the simulation
originally described by H15. We only use the BATS-R-US (Block-Adaptive-Tree-Solarwind-Roe-Upwind-Scheme)
global MHD results here, leaving the other regions covered by SWMF,
such as the ionosphere and ground magnetometer response, to potential
future work. We focus on the dayside and near flanks of the magnetosphere
($X_{GSM}>-15\,\mathrm{R_{E}}$) for which the grid-resolution of
the simulation is $\nicefrac{1}{8}\,\mathrm{R_{E}}$ throughout, apart
from around the inner boundary where a $\nicefrac{1}{16}\,\mathrm{R_{E}}$
resolution shell is used between $2.5\text{\textendash}5\,\mathrm{R_{E}}$
geocentric distance. 

A proxy for the magnetopause location is used within the CCMC tools,
given by the last closed field line along geocentric rays through
a bisection method accurate to $0.01\,\mathrm{R_{E}}$ (fields are
interpolated in the tracing). Throughout, perturbations (represented
by $\delta$'s) from the background (represented by subscript $0$'s)
are defined as the difference to the linear trend. We focus on the
resonant response following the driving phase, i.e. neglecting the
transient wave activity directly driven by the pulse. This is done
by expressing the time that the magnetopause returns to equilibrium
as a function of $X_{GSM}$, extending this throughout the grid, and
then adding half the lowest wave period present in the boundary motion
(A21).  Vector quantities are rotated into local field-aligned coordinates
where the field-aligned direction $\mathbf{e}_{\parallel}$ points
along the time-average of the background magnetic field, the (perpendicular)
azimuthal direction $\mathbf{e}_{\phi_{\perp}}=\left(\mathbf{e}_{\parallel}\times\mathbf{r}\right)/\left|\mathbf{e}_{\parallel}\times\mathbf{r}\right|$
points eastwards ($\mathbf{r}$ is the geocentric position), and the
(perpendicular) radial direction $\mathbf{e}_{r_{\perp}}=\mathbf{e}_{\phi_{\perp}}\times\mathbf{e}_{\parallel}$
is directed outwards (note this exhibits a discontinuity along the
centre of the cusps due to a reversal of direction either side). 

\section{Results}

At each grid point, we compute power spectra of the traces (sums over
all components) of the velocity and magnetic fields respectively.
Spectral peaks whose prominence (how much the peak stands out from
the surrounding baseline) is greater than the two-tailed 95\% confidence
interval of the spectral estimator have been identified. These reveal
spikes in occurrence for both physical quantities at $1.8$, $3.1$,
$6.8$, and $11.7\,\mathrm{mHz}$, with the lowest frequency being
$\sim2\text{\textendash}8$ times more prevalent than the others.
While frequencies above $5\,\mathrm{mHz}$ were not discussed by A21,
the authors showed that the lowest of these frequencies originates
at the subsolar magnetopause as MSE whereas the $3.1\,\mathrm{mHz}$
frequency corresponds to intrinsic Kelvin-Helmholtz waves at the flanks
(peak frequencies vary with local time from $2.5\text{\textendash}3.3\,\mathrm{mHz}$
but only the higher frequencies are prominent). Throughout this paper
we focus on the most widespread frequency of $1.8\,\mathrm{mHz}$,
though we discuss how the results may be generalised to different
frequencies and/or spatial scales.

\subsection{Wave amplitudes and phases\label{subsec:Wave-amplitudes}}

Figure~\ref{fig:maps} shows maps of average power spectral densities
and phases across the $1.0\text{\textendash}2.1\,\mathrm{mHz}$ frequency
band from a Fourier transform of the response phase data. Panels~a--e
show power in the $Z_{GSM}=-2\,\mathrm{R_{E}}$ plane, though other
near-equatorial slices proved similar. We do not show phases here
since there is an ambiguity over which perpendicular directions are
most appropriate. Panels~f--n show results for the noon meridian,
where azimuthal velocities and magnetic fields are zero by symmetry,
with Movie~S1 also showing bandpass filtered results in this plane
(unfiltered movies were presented in A21). To reduce edge effects
in the filtering, the first local maxima/minima in the response phase
is located at each point, with the data being mirrored before this.
A minimum-order infinite impulse response filter with passband range
of $1.0\text{\textendash}2.6\,\mathrm{mHz}$ was applied in both the
forward and reverse directions (zero-phase).

\subsubsection{Near-equatorial planes\label{subsec:Near-equatorial-planes}}

The distribution of Alfv\'{e}n speeds near the equatorial plane,
as shown in Figure~\ref{fig:maps}a, has approximate axial symmetry
at $L$-shells less than $\sim7\,\mathrm{R_{E}}$, where the relative
variation with local time is below $8\%$, however asymmetries rapidly
increase (coefficients of variation up to $\sim50\%$) with radial
distance beyond this point. The simulation does not include a plasmasphere
\citep[cf.][]{claudepierre16}, and thus wave speeds are monotonic
with geocentric distance. The thickness of the magnetopause varies
significantly with local time, evident in the Alfv\'{e}n speed as
a clear enhancement around the last closed field lines (dashed black
line) on the nightside, corresponding well with the current density.
This is $\sim5\,\mathrm{R_{E}}$ thick at $X_{GSM}=-15\,\mathrm{R_{E}}$
though becomes thinner as you go towards the dayside. While determining
the magnetopause thickness from the Alfv\'{e}n speed is less obvious
across the dayside, continuing the locus of points from the nightside
to the subsolar point gives good agreement with the thickness of the
current layer ($9.25\,\mathrm{R_{E}}<X_{GSM}<10.75\,\mathrm{R_{E}}$
as shown by H15).

Velocity and magnetic field perturbations in the near-equatorial magnetosphere
(panels b--e) are generally strongest nearer the magnetopause and
decay with distance from the boundary. There is a clear trend in
power with local time also, being weakest around noon and increasing
as you go further tailward. As noted by A21, this is due to wave growth
of the existing surface modes by the Kelvin-Helmholtz instability,
which can subsequently couple to other modes \citep{southwood74,pu83,kivelson85}.
Parallel velocities (panel~c) are generally stronger than perpendicular
(the total power across both perpendicular components is shown) ones,
indicating wavenumbers $k_{\phi_{\perp}}\ll k_{\parallel}$ which
is generally the case for dayside surface eigenmodes (\citealp{plaschke11};
A21). The opposite scenario, present only near the nightside magnetopause,
is more typical for tailward travelling surface waves or fast magnetosonic
waveguide modes \citep{pu83,mann99}. The dayside response is also
predominantly compressional (panel~e) since the surface eigenmode
is sustained by pressure imbalances across the boundary \citep{plaschke11}.
In contrast, it is the transverse disturbance of the boundary that
is of primary importance in Kelvin-Helmholtz waves \citep{southwood68,hasegawa75},
hence why perpendicular magnetic field perturbations (panel~d) become
larger on the nightside. Finally, nodal structure across the field
is present. $\delta\mathbf{v}_{\perp}$ exhibits subtle antinodes
highlighted in panel~b. These correspond to surface waves (at/near
the magnetopause) or Alfv\'{e}n modes (deeper in the magnetosphere).
$\delta B_{\parallel}$ has several antinodes (peaks) and nodes (troughs)
also, indicative of waveguide modes \citep{waters02}. Their alignment
is not purely radial, as expected in a symmetric setup, instead appearing
to vary with position --- furthest downtail and closest to the magnetopause
they seem to be standing in approximately $\pm Y_{GSM}$, whereas
deeper into the magnetosphere and closer to Earth their normals become
more radially oriented. These agree with the gradients of the reciprocal
Alfv\'{e}n speed and thus the refraction of fast waves \citep{wright18,elsden19}.

\subsubsection{Noon meridian}

\begin{sidewaysfigure*}
\begin{centering}
\noindent \makebox[\textwidth]{\includegraphics[scale=0.9]{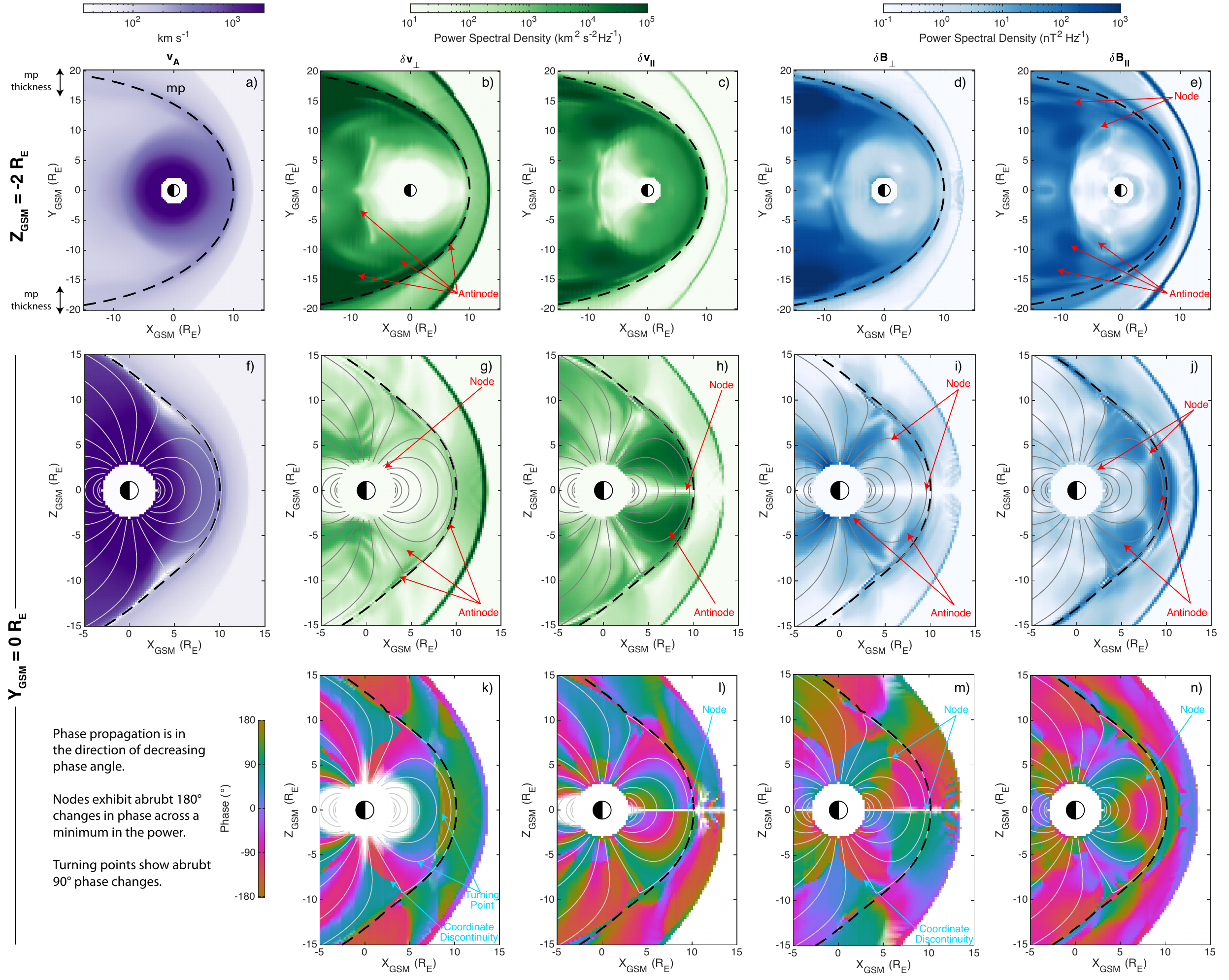}}
\par\end{centering}
\caption{Maps of Alfv\'{e}n speed (a, f), wave power spectral density (b--e,
g--j), and wave phase (k--n) throughout slices of the simulation
(arranged by rows) applied to parallel and both perpendicular components
of the velocity and magnetic field (arranged by columns). The equilibrium
magnetopause location is indicated by the dashed line. Areas of low
power are coloured white in panels k--n.\label{fig:maps}}
\end{sidewaysfigure*}

Figure~\ref{fig:maps}f, showing a cut in the noon meridian, indicates
the simulation magnetic field becomes highly non-dipolar for field
lines with high latitude footpoints, due to the presence of the magnetopause
and cusps. This, along with the accumulation of plasma in the exterior
cusp regions, then affects the Alfv\'{e}n speed map shown --- for
instance there are clear decreases in the cusp regions.

\begin{figure*}
\centering{}\noindent \makebox[\textwidth]{\includegraphics[scale=0.9]{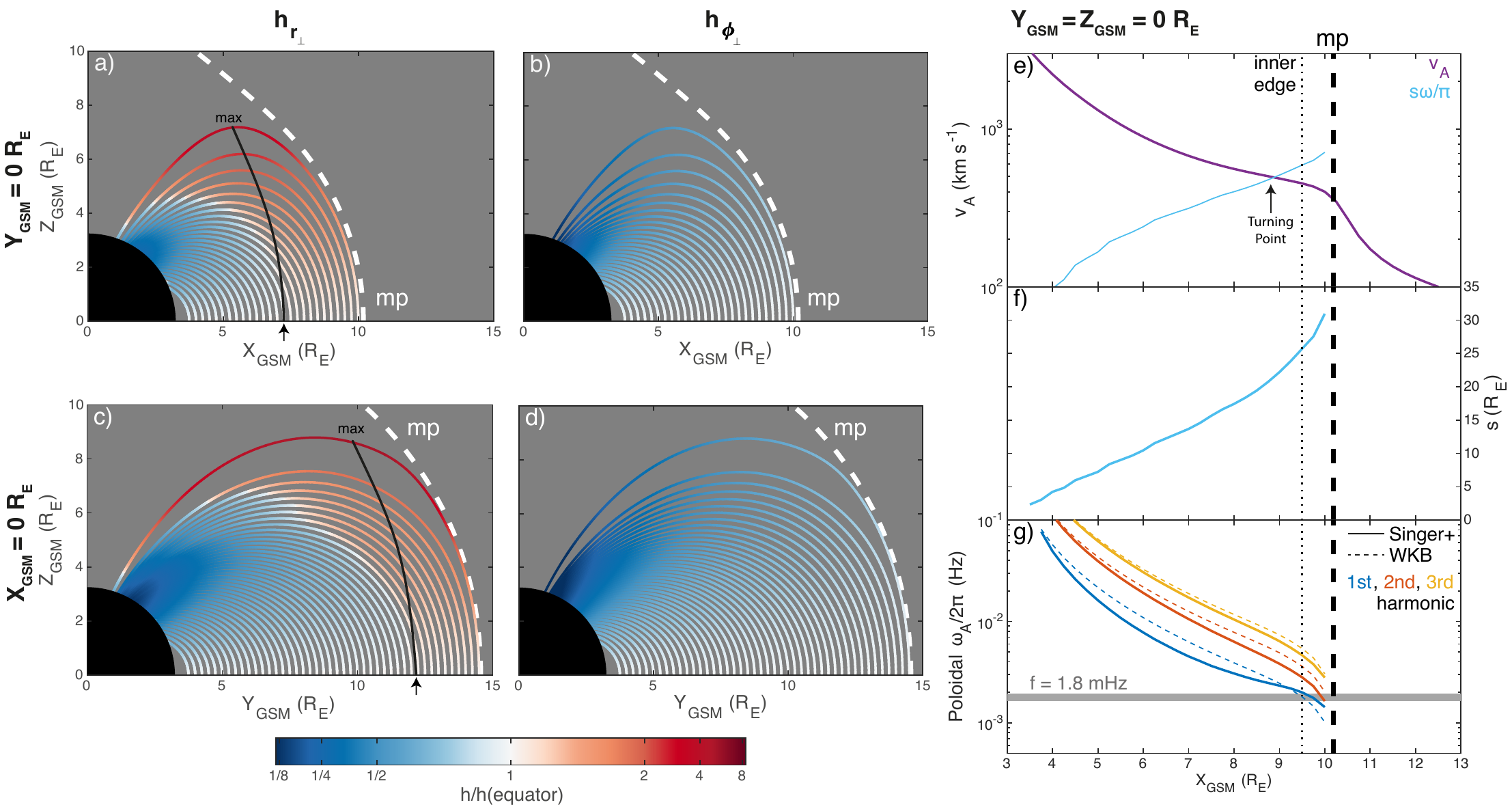}}\caption{Projected field line tracings in the noon (a--b) and dusk (c--d)
meridians coloured by the radial (a, c) and azimuthal (b, d) scale
factors, normalised by their equatorial value. Scale factor local
maxima are indicated by the black line. Also shown are cuts along
the subsolar line of the Alfv\'{e}n speed (e), field line length
(f), and estimated first three harmonics of poloidal Alfv\'{e}n modes
(g) using WKB (dotted) and \citet[solid]{singer91} methods.\label{fig:scalefactor-flr}}
\end{figure*}

The phase of the radial velocity perturbations (Figure~\ref{fig:maps}k)
has a sharp $90^{\circ}$ shift, indicative of a turning point \citep{samson92,rickard94},
Earthward of the magnetopause inner edge (at $X_{GSM}=8.75\,\mathrm{R_{E}}$
on the equator). To understand this theoretically, firstly we trace
field lines along the subsolar line and compute radial scale factors
using the \citet{singer91} method (valid here as little reconnection
present means background field-aligned currents are minimal; \citealp{stern70,salat2000}),
resulting in Figure~\ref{fig:scalefactor-flr}a. From these we can
compute field line lengths $s$ (panel~f) revealing the turning point
occurs when the Alfv\'{e}n speed (panel~e) equals the observed wave
frequency times by twice the field line length (blue line). This is
as expected for a fast mode in cold plasma with zero azimuthal wavenumber
and fundamental standing structure along the field (equation~\ref{eq:magnetosonic-dispersion}).
However, the local frequencies of poloidal Alfv\'{e}n modes in the
magnetosphere (computed using both the WKB and \citealp{singer91},
methods as displayed in panel~g) are higher than that observed. In
fact, the Alfv\'{e}n frequency only becomes as low as $1.8\,\mathrm{mHz}$
on the closed field lines within the magnetopause boundary itself.
\citet{kozyreva19} suggested that resonant coupling between surface
(which is large scale across the field, as observed in the simulation)
and Alfv\'{e}n (which has smaller transverse scales) modes might
occur within the transition layer between magnetosheath and magnetospheric
plasmas, with this coupling potentially providing a means for surface
modes on a boundary of finite thickness to dissipate energy \citep{chen74,lee86,uberoi89}.
Therefore, despite the Alfv\'{e}n speed and frequency profiles being
monotonic with distance from the magnetopause, we find that surface
modes can have turning points which are external to the boundary and
thus the usual expected ordering of turning points and resonance locations
does not always hold in a realistic magnetosphere. The result should
generalise to higher harmonic surface modes since $\omega_{MSE}\propto k_{\parallel}$
(equation~\ref{eq:MSE}). Similar effects were discussed by \citet{southwood86}
in a box model magnetosphere with inhomogeneities along the field
as well as transverse to it, again even if profiles are monotonic.

\begin{figure}
\begin{centering}
\noindent \makebox[\textwidth]{\includegraphics{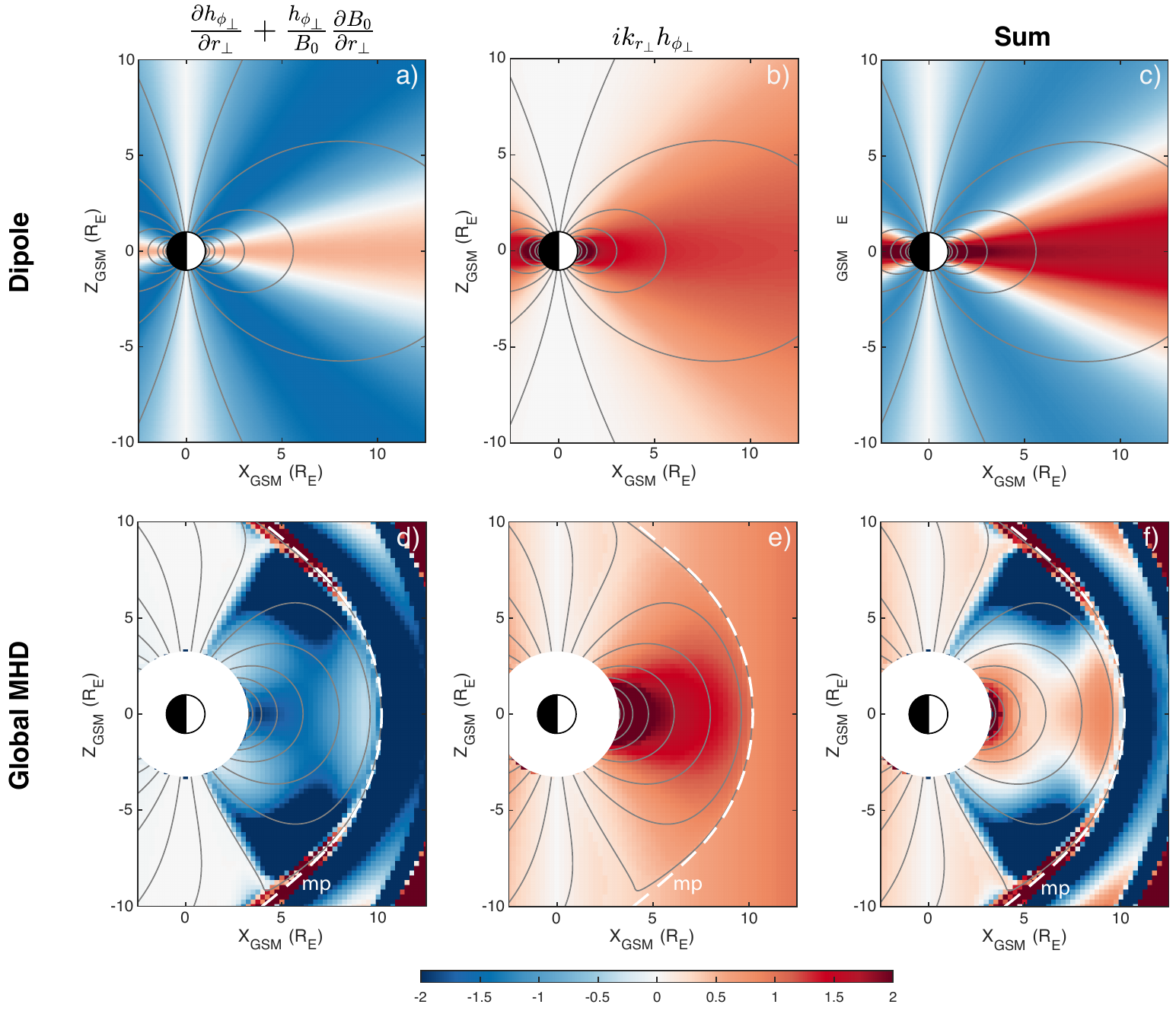}}
\par\end{centering}
\caption{Terms affecting the compressional magnetic field perturbations from
equation~\ref{eq:compression2} dependent on the background field/geometry
(a, d) and the wave (b, e), as well as their sum (c, f). These are
applied to both a dipole field (a--c) and that from the global MHD
simulation (d--f) for the $Y_{GSM}=0\,\mathrm{R_{E}}$ plane.\label{fig:B_par}}
\end{figure}

Abrupt amplitude and phase structure is also present along the field
too. The parallel velocity, shown in panels h and l of Figure~\ref{fig:maps},
has a node (minimum in power and reversal of phase) at the equator
with the phase being relatively constant along field lines either
side. The perpendicular velocity (panel~k), however, is roughly in-phase
all along each field line down to the inner boundary, i.e. there are
no nodes present along the field. Both these points are also evident
in Movie~S1 (middle and right). This standing structure is in agreement
with a fundamental surface eigenmode (Figure~\ref{fig:modes-cartoon}b;
\citealp{plaschke11}). However, one would expect the perpendicular
magnetic field to have antinodes only at the ionospheres with a single
node located at the magnetic equator. Instead, additional nodes can
be seen at the apexes of the outermost field lines near the cusps
(Figure~\ref{fig:maps}m). These can be intuitively understood as
being due to the magnetic field geometry --- any perpendicular radial
displacement that is large-scale (of order $s$) along the field will
not cause deflection of the magnetic field vector here due to the
more rapid geometry changes, hence the location is a node in $\delta\mathbf{B}_{\perp}$.
Similarly, box models predict only one antinode in the compressional
magnetic field, also at the equator, with nodes only at the ionospheres.
Instead it appears that there are additional antinodes at high latitudes
near the cusps that are in antiphase with that at the equator, which
can be seen in both Figure~\ref{fig:maps}j,n and Movie~S1 (left).
This might be expected for either a third harmonic mode or if the
cusps act to bound the surface mode due to the field's curved geometry,
as suggested by \citet{kozyreva19}. However, we know that at this
frequency the velocity exhibits fundamental structure between the
conjugate ionospheres. These compressional features, therefore, might
be a result of the non-uniform background field too, but are less
intuitive to understand. From the MHD induction equation, the parallel
magnetic field is dictated by
\begin{equation}
\begin{array}{ccl}
\frac{\partial\delta B_{\parallel}}{\partial t} & = & -\nabla\cdot\left(B_{0}\mathbf{v}_{\perp}\right)\\
 & = & -\frac{1}{h_{\alpha}h_{\beta}}\left[\frac{\partial}{\partial\alpha}\left(h_{\beta}B_{0}\delta v_{\alpha}\right)+\frac{\partial}{\partial\beta}\left(h_{\alpha}B_{0}\delta v_{\beta}\right)\right]
\end{array}\label{eq:compression}
\end{equation}
where $\beta$ represents a direction perpendicular to both the background
field and $\alpha$. Noting that $\mathbf{v}_{\perp}=\partial\boldsymbol{\xi}_{\perp}/\partial t$,
equation~\ref{eq:compression} may be expressed throughout the noon
meridian, assuming for simplicity a plane wave in the \emph{$r_{\perp}$}
and $\phi_{\perp}$ coordinates, as

\begin{equation}
\begin{array}{cl}
\delta B_{\parallel} & =-\nabla\cdot\left(B_{0}\boldsymbol{\xi}_{\perp}\right)\\
 & \approx-\frac{B_{0}}{h_{r_{\perp}}h_{\phi_{\perp}}}\left[\left(\frac{\partial h_{\phi_{\perp}}}{\partial r_{\perp}}+\frac{h_{\phi_{\perp}}}{B_{0}}\frac{\partial B_{0}}{\partial r_{\perp}}+ik_{r_{\perp}}h_{\phi_{\perp}}\right)\xi_{r_{\perp}}+\left(\frac{\partial h_{r_{\perp}}}{\partial\phi_{\perp}}+\frac{h_{r_{\perp}}}{B_{0}}\frac{\partial B_{0}}{\partial\phi_{\perp}}+ik_{\phi_{\perp}}h_{\phi_{\perp}}\right)\xi_{\phi_{\perp}}\right]\\
 & \approx-\frac{B_{0}}{h_{r_{\perp}}h_{\phi_{\perp}}}\left(\frac{\partial h_{\phi_{\perp}}}{\partial r_{\perp}}+\frac{h_{\phi_{\perp}}}{B_{0}}\frac{\partial B_{0}}{\partial r_{\perp}}+ik_{r_{\perp}}h_{\phi_{\perp}}\right)\xi_{r_{\perp}}
\end{array}\label{eq:compression2}
\end{equation}
since in our simulation we have $\xi_{\phi_{\perp}}=0$ by symmetry
here. The first two terms depend only on the background field and
its geometry whereas the third term is largely dictated by the wave
itself. The first two terms can be evaluated analytically for a dipole
field, as detailed in \ref{sec:dipole}, to give Figure~\ref{fig:B_par}a.
For the MHD simulation we use the scale factors from the \citet{singer91}
method shown in Figure~\ref{fig:scalefactor-flr}a--b along with
the fact that $\partial a/\partial r_{\perp}=h_{r_{\perp}}\mathbf{e}_{r_{\perp}}\cdot\left(\nabla a\right)$
for derivatives to arrive at Figure~\ref{fig:B_par}d. The radial
wavenumber is taken to be $k_{r_{\perp}}\approx-ik_{\parallel}\approx-i\pi/s$
in panels~b and e, which is true for a fundamental surface eigenmode
under the assumptions of no damping and incompressibility \citep{plaschke11}
(for open field lines in the simulation we keep $s$ fixed as that
of the last closed field line). From the sum of all three terms in
equation~\ref{eq:compression2}, both dipole (panel~c) and MHD (panel~f)
fields predict in the high latitude magnetosphere a reversal in sign
of the proportionality constant between the compressional magnetic
field and the perpendicular displacement. Thus a fundamental mode
yields additional nodes and antinodes in the compressional magnetic
field. In fact, for the MHD simulation, along the outermost closed
field lines (where plasma displacements are largest) there is excellent
agreement in the patterns present in the observed compressional perturbations
(Figure~\ref{fig:maps}j,n) and the predictions based on equaton~\ref{eq:compression2}
(Figure~\ref{fig:B_par}f). Therefore, a realistic magnetic field
can introduce additional structure to the compressional magnetic field
oscillations associated with surface modes that are not predicted
by box models. As the lowest frequency dayside normal modes of the
magnetosphere though, surface eigenmodes have the smallest radial
wavenumbers and, following equation~\ref{eq:compression2}, are perhaps
most affected by geometrical effects. It is therefore instructive
to consider back-of-the-envelope calculations for the other MHD wavemodes.
We predict, given the values shown in Figure~\ref{fig:B_par}d, that
most cavity/waveguide modes and some poloidal Alfv\'{e}n waves should
be altered by the non-uniform magnetic field --- only those with
short radial extents ($k_{r_{\perp}}^{-1}\ll1\text{\textendash}2\,\mathrm{R_{E}}$)
ought to be relatively unaffected. We leave testing these to future
work.

Figure~\ref{fig:maps}k--n indicates, via the gradual decreases
in phase radially, the slow perpendicular phase motion discussed by
A21 as a result of the surface mode damping. This is even more evident
in Movie~S1. The movie also reveals interesting behaviour at the
magnetospheric cusps. As the outermost traced magnetospheric field
line shown is displaced by the surface mode, as indicated by the radial
velocity, these perturbations clearly propagate through the cusps
away from the magnetosphere. Rather than being evanescent, the disturbances
appear not to decay in amplitude with distance here, which is backed
up by the presence of another turning point at the location of the
outermost magnetospheric field line as indicated in Figure~\ref{fig:maps}k.
Similar propagating behaviour is also seen for the compressional magnetic
field at the cusps in Movie~S1. While surface modes are often treated
theoretically under the assumption of incompressibility, which gives
evanescent behaviour on both sides of the boundary, the plasma in
the exterior cusps is similar to that in the magnetosheath and thus
highly compressible \citep{archer15}. This fact predicts (via equation~\ref{eq:magnetosonic-dispersion})
propagating rather than evanescent magnetosonic waves, as mentioned
by A21 in explaining why the subsolar bow shock motion lags the magnetopause
by the fast magnetosonic travel time.

\subsection{Polarization ellipses}

We now investigate the polarizations of the oscillations present throughout
the magnetosphere. The orientation and ellipticity parameters of the
polarization ellipse as well as the degree of polarization are calculated
as detailed in \ref{sec:Polarisation-parameters} for the perpendicular
perturbations in the velocity and magnetic field.

\subsubsection{Velocity polarization}

\begin{figure*}
\begin{centering}
\noindent \makebox[\textwidth]{\includegraphics{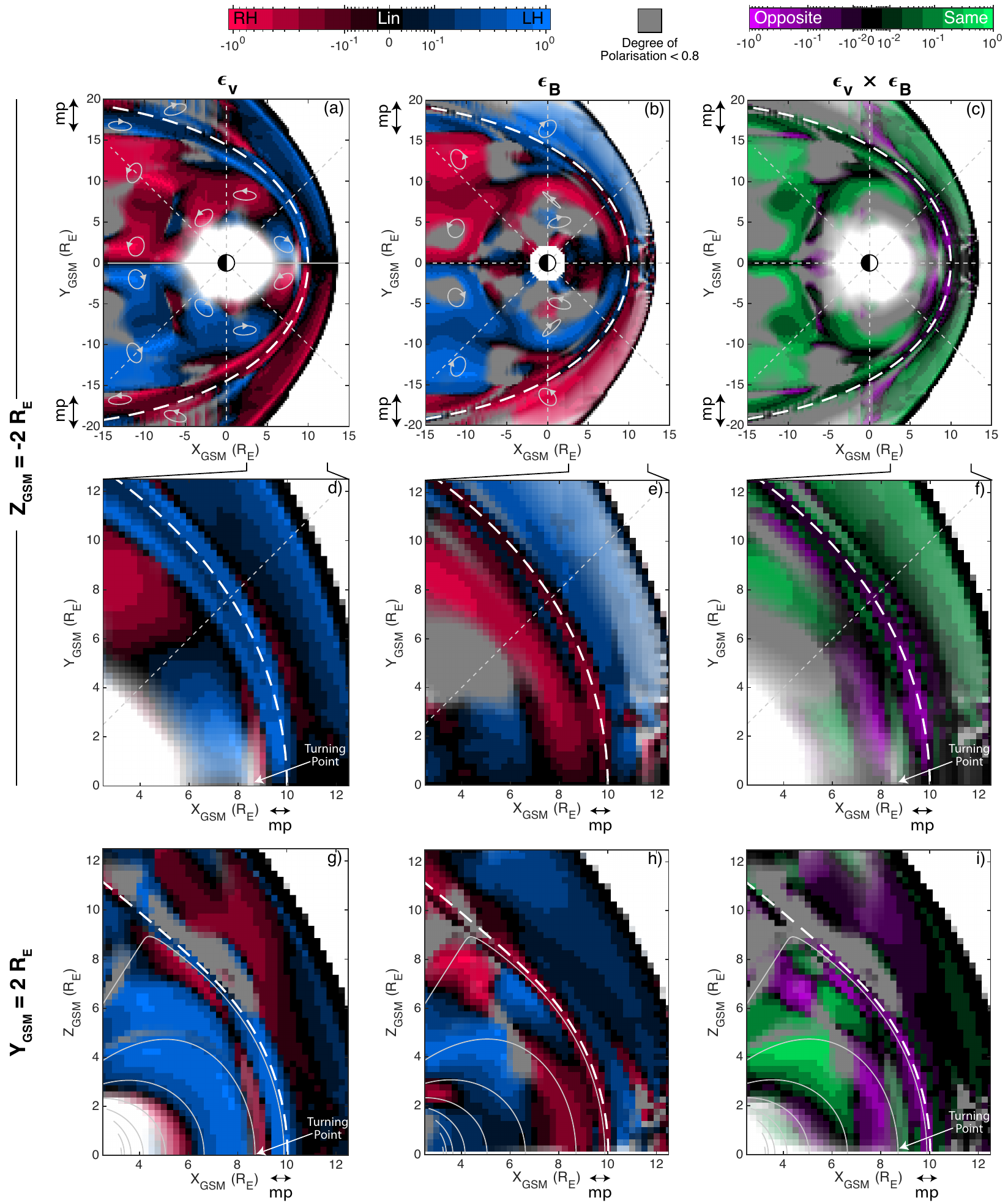}}
\par\end{centering}
\caption{Polarization ellipticities in simulation slices shown for the velocity
(a, d, g) and magnetic field (b, e, h). Their product is also shown
(c, f, i). Colour scales use a bi-symmetric log transform \citep{webber12}.
Regions with low degree of polarization and power are coloured grey
and white respectively.\label{fig:handedness}}
\end{figure*}

\begin{figure*}
\begin{centering}
\noindent \makebox[\textwidth]{\includegraphics{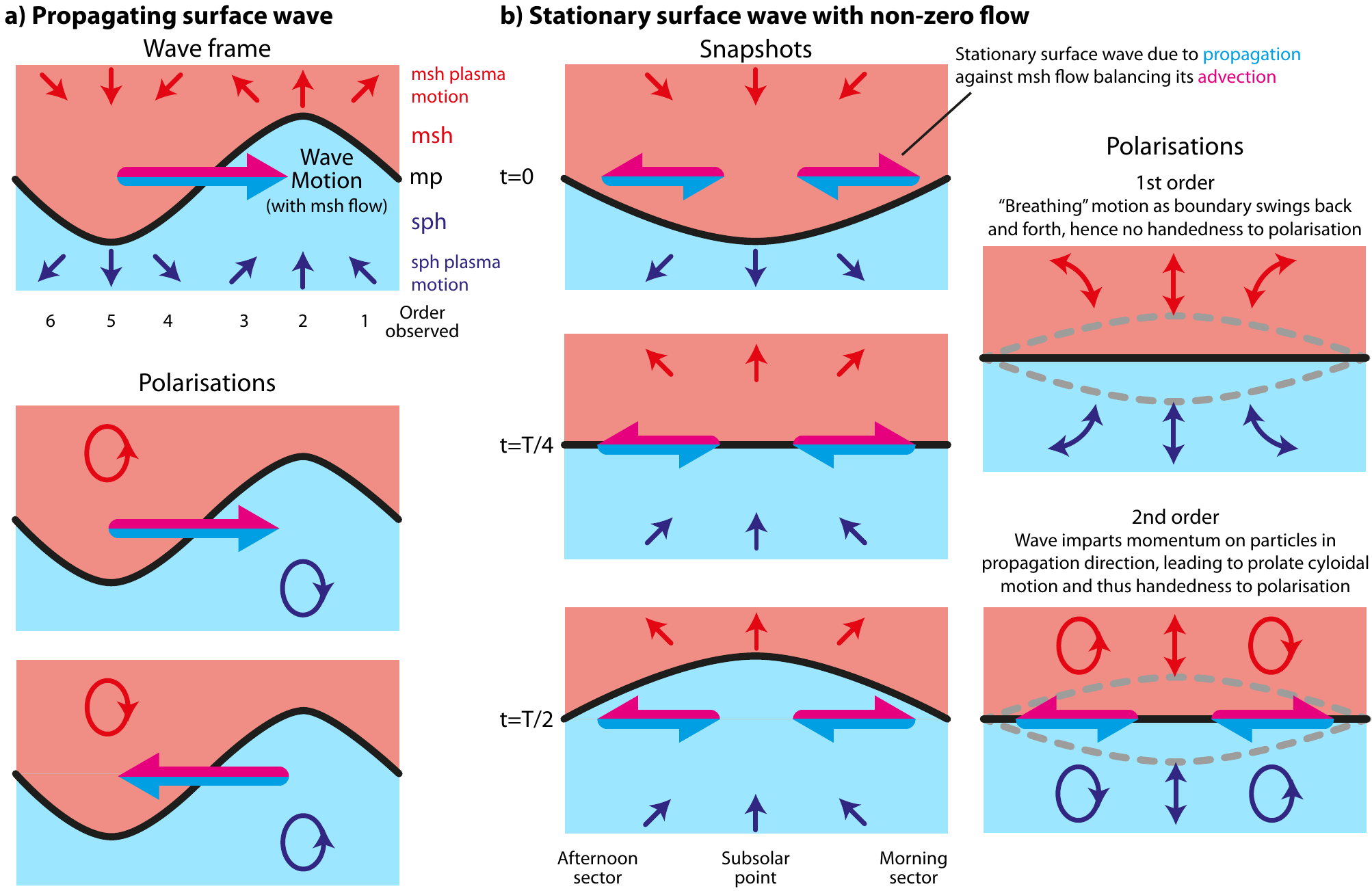}}
\par\end{centering}
\caption{Illustration of the expected velocity polarizations for a propagating
(a) and stationary (b) magnetopause surface wave. Plasma motion arrows
indicate the displacement that parcel undertook in the last quarter
cycle ($T/4$). \label{fig:polarisation-cartoon}}
\end{figure*}

Figure~\ref{fig:handedness}a shows ellipticities of the velocity
in the $Z_{GSM}=-2\,\mathrm{R_{E}}$ plane. These are antisymmetric
about the noon--midnight meridian, with opposite handedness either
side, due to the symmetry of the simulation. Throughout the magnetosphere
for local times before $09h$ and after $15h$ the polarizations are
largely left- and right-handed with respect to the magnetic field
respectively. A21 showed that the magnetopause perturbations propagate
tailward at these local times. The results agree with expectations
for this scenario, as illustrated in Figure~\ref{fig:polarisation-cartoon}a
by showing the plasma displacement either side of the boundary in
the frame of a surface wave (top) and how this results in a sense
of rotation in the Earth's frame as the wave propagates tailward with
the magnetosheath flow (bottom) \citep{stokes1847,southwood68,samson71}.
Either side of noon, the perturbations within the magnetopause current
layer are right-handed on the dawn-side and left-handed on the dusk-side.
These are consistent with the \citet{lee81} model of surface waves
in a boundary layer of finite thickness, where the polarization inside
the transition layer is dominated by the mode at the interface with
the magnetosheath rather than that with the magnetosphere.

Between $09h$ and $15h$ MLT it was shown by A21 that the magnetopause
is a stationary wave, with this being achieved by the surface waves
propagating against and pefectly balancing the tailward magnetosheath
flow. Figure~\ref{fig:handedness}d shows a zoom in of the dayside
post-noon sector. This reveals between noon and $15h$ local time
that the handedness of the velocity perturbations are mostly left-handed
with respect to the magnetic field, the opposite of that found later
in the afternoon. An exception occurs between the inner edge of the
magnetopause and the turning point identified previously, where right-handed
waves are present. Figure~\ref{fig:handedness}g shows these polarizations
remain consistent along the field also. To understand the polarization
Earthward of the turning point, where the waves are evanescent, we
show snapshots in time of a stationary surface wave in Figure~\ref{fig:polarisation-cartoon}b.
To first order, a stationary surface wave has no handedness to its
polarization, as shown in the top right image of Figure~\ref{fig:polarisation-cartoon}b,
since the boundary undergoes a simple breathing motion \citep[e.g.][]{lamb32}.
However, this prediction neglects any background flow or wave propagation.
\citet{stokes1847} showed by taking into account the evanescent nature
of a travelling surface wave's flow patterns, fluid elements' paths
are no longer perfect orbits but become cycloidal having moved greater
distance in the direction of propagation closer to the boundary than
in the opposite direction when farther away (see also \citealp{southwood93}).
Therefore, a surface wave imparts momentum on the particles in the
direction of propagation. Since in a stationary magnetopause surface
wave subject to non-zero magnetosheath flow the wave propagates towards
the subsolar point (A21), momentum will thus be imparted on the plasma
in the same sense. This results in a handedness to the polarization,
as depicted in bottom right panel of Figure~\ref{fig:polarisation-cartoon}b,
which is in agreement with the simulation. The handedness of velocity
perturbations in the magnetosphere could thus be used to infer stationary
surface waves in spacecraft observations. Between the turning point
and the magnetopause the waves are not evanescent, thus a reversal
in polarization is expected \citep{southwood74}.

Since \citet{samson71}, observations have predominantly reported
a reversal of ULF wave polarizations only around noon \citep[e.g.][]{ziesolleck95,mathie99},
conforming with the expectations of tailward propagating disturbances.
The polarization patterns presented here thus appear to differ to
the typical pattern, though we note other exceptions are reported
in the literature. Of note is the recent work of \citet{huang21},
who present a case study of global $10\,\mathrm{min}$ period waves
observed by $\sim180$ ground magnetometers following a solar wind
presssure pulse that are consistent with our simulation results. The
author found that the waves (whose period was independent of latitude/longitude
and amplitudes increased with latitude) had rotating equivalent currents
in the northern hemisphere which were clockwise (i.e. right-handed
with respect to the field) in the morning and evening sectors, and
anticlockwise (i.e. left-handed with respect to the field) in the
post-midnight and afternoon sectors --- in agreement with the polarizations
presented here. Based on our results, we therefore interpret these
observations as due to MSE excited by the impulse.

The projected orientation of the velocity polarization ellipses are
also shown in a few locations in Figure~\ref{fig:handedness}a. Near
the Earth, the semi-major axes are aligned predominantly in the azimuthal
direction, though there is a non-negligible radial component also
(e.g. around noon the polarisation is almost entirely radial) . In
the nightside ($X_{GSM}<-5\,\mathrm{R_{E}}$) the ellipses tend to
be aligned more to lines of constant $X_{GSM}$. The simulation, therefore,
reproduces the fact that a realistic magnetosphere changes the directions
of MHD waves' velocity (or equivalently electric field) oscillations
compared to those predicted by models with perfect cylindrical symmetry
\citep{rankin06,kabin07,degeling14,wright20}. Despite the waves present
being predominantly compressional, these orientations are approximately
perpendicular to the gradient in amplitude as expected for non-compressional
modes rather than parallel to it \citep{southwood84}. In spacecraft
observations this has been regularly observed and interpreted as evidence
of (non-resonant) wave coupling between fast and Alfv\'{e}nic modes.

\subsubsection{Magnetic field polarization}

The same polarization analysis is shown for the magnetic field perturbations
in Figure~\ref{fig:handedness}b, e, and h. In a uniform background
field, Alfv\'{e}n's frozen-in theorem predicts these should have
the same sense as the velocity. Therefore in panels~c, f, and i we
also show the product of the two ellipticities, which indicate regions
where they are the same (green) or opposite (purple). Throughout most
of the near-equatorial slice they indeed have the same handedness.
However, the zoom in on the dayside (panel~f) highlights a sizeable
region within the magnetosphere where their polarizations are opposite
(while there are some other instances, these mostly occur within boundary
layers or the cusps). Along the subsolar line this region extends
from the turning point to $X_{GSM}=7.5\,\mathrm{R_{E}}$. Panel~i
shows that this opposite polarization doesn't extend all the way along
the field lines, terminating at some point that extends further in
$Z_{GSM}$ for larger $L$-shells. This means that near the inner
boundary of the simulation the magnetic field returns to having the
same handedness as the velocity. 

To the best of our knowledge, opposite handedness in the polarizations
of velocity and magnetic field oscillations has not been reported
before. It is likely that this is due to the non-uniform background
field. \citet{singer91} show that the magnetic perturbations in an
arbitrary orthogonal field geometry are related to the displacement
via
\begin{equation}
\delta B_{\alpha}=h_{\alpha}B_{0}\frac{\partial}{\partial s}\left(\frac{\xi_{\alpha}}{h_{\alpha}}\right)\label{eq:singer-dB}
\end{equation}
, valid here due to little shearing being present \citep{stern70,salat2000}.
In Figure~\ref{fig:scalefactor-flr}a--b we show how the radial
and azimuthal scale factors vary along each field line in the noon
meridian. It is clear that the azimuthal scale factors decrease in
value along the field either side of the equator. This is also true
of the radial scale factors for low $L$-shells, as expected for an
approximately dipolar field (see \ref{sec:dipole}). However, from
$X_{GSM}\geq7.5\,\mathrm{R_{E}}$ (indicated by the black arrow) $h_{r_{\perp}}$
increases with distance along the field from the equator reaching
a maximum as field lines become further apart towards the cusps. The
locus of these local maxima are depicted by the black line in panel~a.
$h_{r_{\perp}}$ changes by up to $4\text{\textendash}5\times$ its
equatorial value over a fraction of the field line length ($\sim5\text{\textendash}30\%$),
whereas the scale length along the field of the displacement is $2s$
(a fundamental mode). Therefore, from equation~\ref{eq:singer-dB},
one would expect geometric effects to dominate as we are in the long
wavelength limit. Since $\partial h/\partial s$ has opposite signs
for the radial and azimuthal directions in the region to the right
of the black line, this should lead to a reversal in handedness of
the magnetic polarisation compared to the displacement (and thus also
the velocity). This region agrees extremely well with the polarizations
observed in the simulation, with Figure~\ref{fig:handedness}i clearly
showing a matching trend with $Z_{GSM}$. We look for similar evidence
of a reversal of handedness along the terminator also, with Figure~\ref{fig:scalefactor-flr}c--d
showing the traced field lines and scale factors. These predict a
reversal at $Y_{GSM}=\pm12.25\,\mathrm{R_{E}}$. Figure~\ref{fig:handedness}b
does indeed show a reversal in the handedness of the magnetic field
around these points. Unfortunately though the velocity perturbations
in this region are almost linearly polarized and thus it is unclear
whether the magnetic field and velocity are of opposite polarization
here. Nonetheless, our results highlight that care needs to be taken
when using the polarization of the magnetic field from spacecraft
observations \citep[e.g.][]{takahashi91,kokubun13,agapitov09}, as
close to the magnetopause this can be reversed with respect to the
displacement/velocity purely due to the highly curvilinear geometry
present and likely affects many developed ULF wave diagnostics based
on simple models. This is likely the case also at the other planetary
magnetospheres, where ULF waves have been studied but often only magnetic
field measurements are available \citep[e.g.][]{james16,manners19}.
These effects, however, do not appear to influence terrestrial magnetic
perturbations measured from the ground since $\partial h/\partial s$
has the same sign in both directions close to the Earth, and thus
previous results from networks of ground magnetometers \citep[e.g.][]{samson71,ziesolleck95,mathie99}
remain reliable.

The example polarization ellipses shown in Figure~\ref{fig:handedness}b
also indicate that the orientation of magnetic perturbations can differ
to those in the velocity too. While the two are somewhat similarly
oriented on the nightside, we find that on the dayside the magnetic
field semi-major axes tend be predominantly radial in orientation,
differing from the velocity by $\sim50\text{\textendash}90^{\circ}$
(apart from around noon where both quantities are radially aligned).
The fact that the magnetic field's polarization can have a different
orientation has not been stressed in the previous literature, since
such studies have largely focused on either the displacement \citep{singer91}
or electric field \citep{rankin06,kabin07,degeling14}.

\subsection{Standing wave detection for spacecraft}

Section~\ref{subsec:Wave-amplitudes} detailed the presence of standing
structure spatially within the simulation, both across the geomagnetic
field and along it. Outside of a simulation though, it is generally
not possible to infer this due to spacecraft observations being sparse
and subject to spatio-temporal ambiguity. One common method of detecting
standing waves is derived from the wave Poynting flux
\begin{equation}
\begin{array}{cl}
\mathbf{S} & =\delta\mathbf{E}\times\delta\mathbf{B}/\mu_{0}\\
 & =\left(\mathbf{B}_{0}\times\delta\mathbf{v}\right)\times\delta\mathbf{B}/\mu_{0}\\
 & =\left[-\left(\delta\mathbf{B}_{\perp}\cdot\delta\mathbf{v}_{\perp}\right)\mathbf{B}_{0}+B_{0}\delta B_{\parallel}\delta\mathbf{v}_{\perp}\right]/\mu_{0}
\end{array}\label{eq:poynting-inst}
\end{equation}
and requiring the net energy propagation averaged over a cycle be
zero in the direction the wave is standing \citep{kokubun77}. Using
phasor notation, where instantaneous fields go as $\delta\mathbf{v}\left(t\right)=\delta\tilde{\mathbf{v}}e^{i\omega t}$
with the tilde indicating the phasor (complex amplitude), the complex
Poynting vector can be constructed as

\begin{equation}
\begin{array}{cl}
\tilde{\mathbf{S}} & =\delta\tilde{\mathbf{E}}\times\delta\tilde{\mathbf{B}}^{*}/2\mu_{0}\\
 & =\left[-\left(\delta\tilde{\mathbf{B}}_{\perp}^{*}\cdot\delta\tilde{\mathbf{v}}_{\perp}\right)\mathbf{B}_{0}+B_{0}\delta\tilde{B}_{\parallel}^{*}\delta\tilde{\mathbf{v}}_{\perp}\right]/2\mu_{0}
\end{array}\label{eq:poynting-phasor}
\end{equation}
The time-averaged power flow is then simply given by $\mathrm{Re}\left\{ \tilde{\mathbf{S}}\right\} $.
In contrast, $\mathrm{Im}\left\{ \tilde{\mathbf{S}}\right\} $ corresponds
to reactive power --- the flow of trapped energy that converts between
electric and magnetic components without contributing to the propagation
of the field --- which can indicate the presence of standing waves.
From equation~\ref{eq:poynting-phasor} we see that standing waves
either parallel or perpendicular to the field have $\pm90^{\circ}$
cross-phases between components of the velocity and magnetic fields.
Methods for detecting standing waves in spacecraft data thus search
for this desired cross-phase, though often make assumptions (based
on simplified models) about the orientation of the wave perturbations.

\subsubsection{Standing structure across the field}

Nodal structure across the field on the nightside, which indicates
the presence of waveguide modes, was commented on in section~\ref{subsec:Near-equatorial-planes}.
\citet{waters02} suggested such modes could be found where the compressional
magnetic field and azimuthal electric field (equivalent to radial
velocity) are in quadrature. This assumes the fast magnetosonic waves
are standing in the radial direction, which would have been the case
in the 3D wave simulation used by the authors (that of \citealp{lee99})
since this has a dipole magnetic field and axially symmetric Alfv\'{e}n
speeds. This criterion has been used to detect such modes around noon
in global MHD simulations \citep{hartinger14} as well as in spacecraft
observations \citep{hartinger12,hartinger13}. However, we find in
our simulation that most of the regions that show clear evidence of
monochromatic cavity/waveguide modes do not exhibit this required
phase difference. This is likely because in a realistic magnetosphere
fast waves do not necessarily interfere in the purely radial direction,
though it is not clear which is the most physically appropriate direction
to consider.

\citet{wright20} posed a different diagnostic for standing or propagating
fast waves by explicitly calculating the right-hand side of equation~\ref{eq:compression}
in their dipole field simulations with more realistic (non-axially
symmetric) Alfv\'{e}n speeds. They showed through visual inspection
that along a path parallel to the magnetopause, this was in-phase
with the compressional magnetic field around noon (indicating standing
waves), become more ambiguous, and then was in quadrature at the distant
flank (propagating waves). This method, however, cannot be applied
to spacecraft data since it requires multipoint measurements for the
calculation of derivatives as well as knowledge of the magnetic field
geometry.

Instead, we seek to generalise the \citet{waters02} criterion. Rather
than only using the radial direction, we consider all directions perpendicular
to the background field. For each one we calculate the cross-phase
between that component of $\delta\mathbf{v}_{\perp}$ and the compressional
magnetic field as well as their coherence. We then find in which directions
the desired cross-phase holds to within $\pm22.5^{\circ}$ ($45^{\circ}$
bins centred on the target), is coherent ($>0.8$), and has significant
wave power. The results are depicted in Figure~\ref{fig:standing-directions}a--b
for standing (quadrature) and propagating (in-phase) waves respectively.
In these panels, coloured regions indicate the desired cross-phase
is present in some direction, whereas blacks show this did not occur
and greys depict a lack of coherence. The markers indicate the directions
which satisfied all our criteria.

Figure~\ref{fig:standing-directions}a reveals a number of regions
with standing structure across the magnetospheric magnetic field.
The first of these is within the magnetopause current layer itself,
spanning most of the dayside. Here the components of the velocity
nearly tangent to the boundary are in quadrature with the compressional
magnetic field. The second standing region is in the outer magnetospheric
flanks ($X_{GSM}<-5\,\mathrm{R_{E}}$), corresponding well with the
location of the nodal structure in $\delta B_{\parallel}$ identified
previously. While far into the tail the standing directions are approximately
aligned with $\pm Y_{GSM}$ it is clear that nearer Earth the standing
axis tilts towards the Sun-Earth line. The Earthward edge of the standing
region approximately follows the contour of the Alfv\'{e}n speed
(Figure~\ref{fig:maps}a), suggesting this constrains the penetration
of fast magnetosonic waves from the magnetopause, in line with simple
theory \citep{kivelson84,kivelson85}. However, standing behaviour
is also present within the region of high Alfv\'{e}n speed near the
Earth (centred at $X_{GSM}=-2\,\mathrm{R_{E}}$ and $Y_{GSM}=8\,\mathrm{R_{E}}$)
which does not extend to the magnetopause. Curiously the standing
direction changes considerably within this region, varying from quasi-radial
to almost azimuthal. Such behaviour is not anticipated from models
with axial symmetry and thus is likely due to wave refraction in the
more realistic wave speed profile \citep{wright18,elsden19}. Finally,
we note that within the dayside magnetosphere, where the surface mode
dominates, quadrature emerges only deeper into the magnetosphere.
This corresponds to where the effects of plasma compressibility become
more negligible due to higher Alfv\'{e}n speeds (equation~\ref{eq:magnetosonic-dispersion})
and thus more evanescent and less propagating behaviour is expected,
as mentioned by A21.

We compare these results to Figure~\ref{fig:standing-directions}b,
which indicates where and in which direction waves propagating across
the field are present. Note that colours here are less intense since
there is only one target cross-phase value, whereas for standing waves
there were two. In most regions, the direction of propagation is almost
perpendicular to that in panel~a, i.e. the waves propagate transverse
to the direction in which they are standing. The direction of propagation
throughout is largely in line with the time-averaged Poynting vector
presented by A21 --- towards the subsolar point on the dayside and
generally tailward on the nightside.

While we could have simply used the reactive power component of the
complex Poynting vector, we note this requires considerable care.
Firstly, $\mathrm{Im}\left\{ \tilde{\mathbf{S}}_{\perp}\right\} $
will only yield one direction, thus in cases where waves are standing
in several directions $\mathrm{Im}\left\{ \tilde{\mathbf{S}}_{\perp}\right\} $
will vary with location (true of even simple examples of a homogeneous
rectangular cavity) thus does not always indicate the direction in
which a wave is standing. Secondly, $\mathrm{Im}\left\{ \tilde{\mathbf{S}}_{\perp}\right\} $
can have significant components along or anti-parallel to the time-averaged
Poynting vector. These are clearly unrelated to a purely standing
wave and may be the result of either the interference of waves of
different amplitudes, near-field antenna effects close to a current
source, or interactions between the wave fields and the plasma. Finally,
while $\mathrm{Im}\left\{ \tilde{\mathbf{S}}_{\perp}\right\} $ can
always be computed, it is necessary to determine whether the resulting
vector (or a particular component) is statistically significant and
thus meaningful, especially in the presence of background noise.

\subsubsection{Standing structure along the field}

\begin{figure*}
\begin{centering}
\noindent \makebox[\textwidth]{\includegraphics{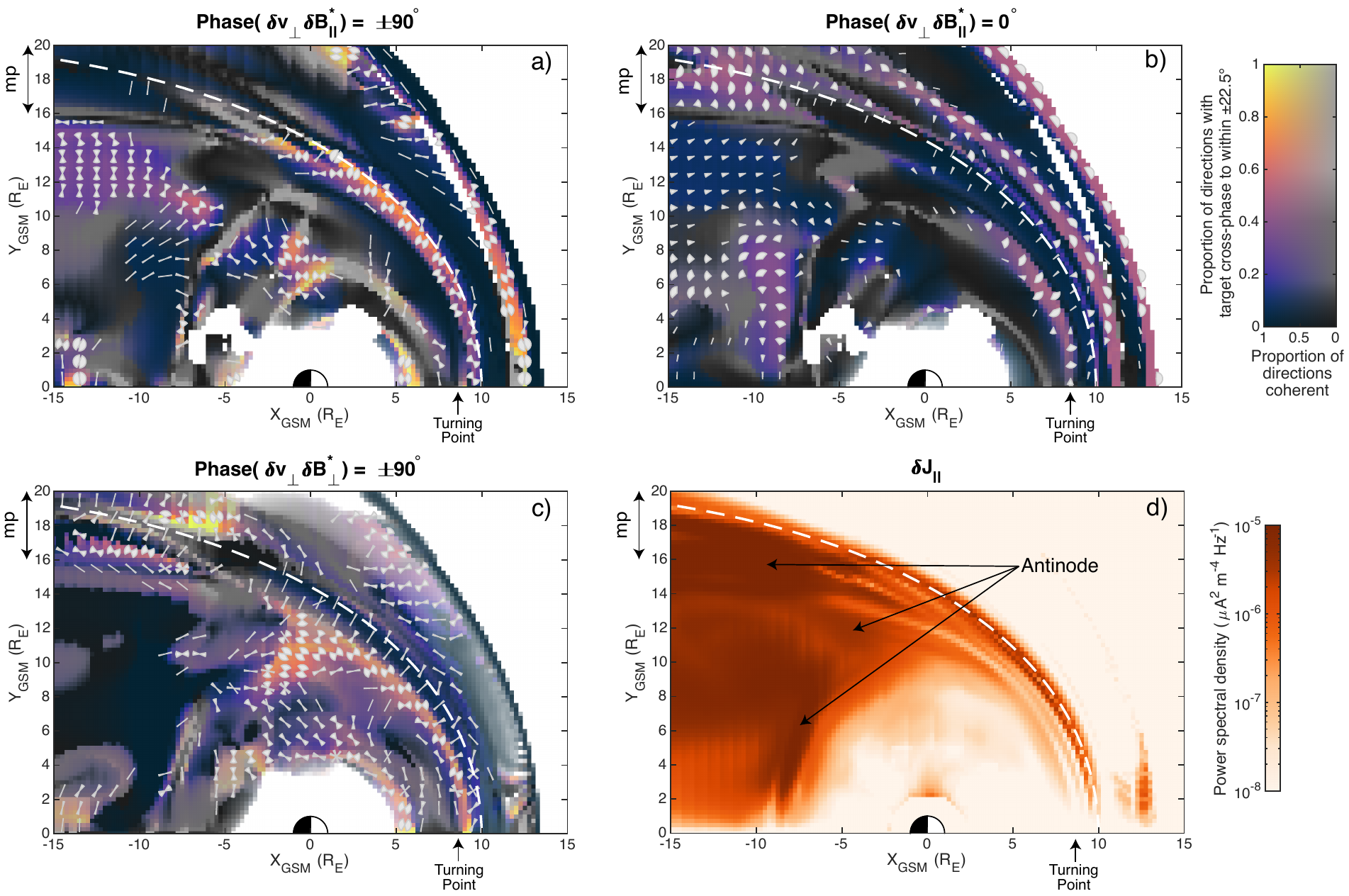}}
\par\end{centering}
\caption{Projection in the $Z_{GSM}=-2\,\mathrm{R_{E}}$ plane of directions
perpendicular to the magnetic field (grey markers) with phase relations
relevant for a) standing structure across the field, b) propagating
structure across the field, and c) standing structure along the field.
Colours show the proportion of directions the phase relation is satisfied,
with greys and whites indicating low coherence and power respectively.
Wave power spectral density in the field-aligned current is also shown
(d).\label{fig:standing-directions}}
\end{figure*}

To detect standing structure along the field, a criterion often used
is a $\pm90^{\circ}$ phase difference between a component of $\delta\mathbf{B}_{\perp}$
and the same component of $\delta\mathbf{v}_{\perp}$ (or equivalently
the component of $\delta\mathbf{E}_{\perp}$ perpendicular to both
that direction and the background magnetic field) \citep[e.g.][]{kokubun77,takahashi84}.
This is generally referred to as a test for standing Alfv\'{e}n waves
and typically the radial or azimuthal directions are chosen. However,
similarly to before, we generalise this criterion to consider all
directions perpendicular to the background field in search of standing
structure along it. The results are shown in Figure~\ref{fig:standing-directions}c
in the same format as before.

Across most of the dayside magnetosphere evidence of standing structure
along the field is present, in line with the results presented earlier.
This is found mostly in the azimuthal direction, likely because the
geometric effects pertaining to the radial direction lead to a different
phase relationship than that predicted within a box or symmetric dipole
model. However, we know that the signatures on the dayside cannot
be explained as Alfv\'{e}n waves because the observed frequency is
too low and the mode of oscillation is primarily compressional. Additional
evidence against a pure Alfv\'{e}n mode is also given in Figure~\ref{fig:standing-directions}d,
which shows a power map (similar to those in Figure~\ref{fig:maps})
for the field-aligned current $\delta J_{\parallel}$. While Alfv\'{e}n
waves are associated with such currents, fast magnetosonic waves (either
propagating or evanescent) are generally not \citep[e.g.][]{wright20}.
Indeed, $\delta J_{\parallel}$ is weak throughout the dayside magnetosphere.
Therefore, we stress that a $\pm90^{\circ}$ phase difference between
$\delta\mathbf{B}_{\perp}$ and $\delta\mathbf{v}_{\perp}$ does not
necessarily indicate a standing Alfv\'{e}n wave, but simply standing
structure along the field. There are, however, strong field-aligned
currents within the magnetopause boundary layer. This agrees with
theoretical predictions for surface eigenmodes in a box model, which
are supported by currents flowing entirely within the boundary that
are closed via the ionospheres \citep{plaschke11}.

On the nightside, in general there is little standing structure along
the field present as evident in Figure~\ref{fig:standing-directions}c.
This may be due to the nightside field lines being much longer, and
thus having eigenfrequencies much lower than $1.8\,\mathrm{mHz}$,
or indeed some field lines not even being closed within the full simulation
domain. We do, however, observe a few areas of localised standing
structure present, e.g. at around $X_{GSM}=-8\,\mathrm{R_{E}}$ near
the central magnetotail as well as along $Y_{GSM}=11\,\mathrm{R_{E}}$
near the terminator. These regions correspond well with the subtle
antinodes in $\delta\mathbf{v}_{\perp}$ identified in Figure~\ref{fig:maps}b
and also exhibit enhanced field-aligned currents, as seen in Figure~\ref{fig:standing-directions}d.
Therefore, we conclude that these do in fact correspond to Alfv\'{e}n
modes.

Similarly to across the field, we find that if only the usual radial/azimuthal
directions are used then most areas with standing structure along
the field are missed. This again highlights the need to consider all
directions perpendicular to the field when in a realistic magnetosphere.
We also note that an alternative test requiring significant reactive
power along the field, $\mathrm{Im}\left\{ \tilde{\mathbf{S}}_{\parallel}\right\} $,
could be used though the same care to that outlined earlier is required.

\section{Summary}

We have investigated the polarizations of system-scale magnetohydrodynamic
(MHD) waves in a realistic magnetosphere through using a global MHD
simulation of the resonant response to a solar wind pressure pulse.
While many aspects of the surface, fast magnetosonic, and Alfv\'{e}n
waves excited are in agreement with simple box models, we find that
some of the predictions are significantly altered. The key findings
are the following:
\begin{enumerate}
\item The often assumed order from the fast magnetosonic dispersion relation
of a turning point followed by matching Alfv\'{e}n resonant location
does not always hold for radially monotonic Alfv\'{e}n speed profiles.
This prediction arises from box models with inhomogeneity only in
the radial direction. \citet{southwood86} showed introducing additional
inhomogeneity along the field within such a model allows fast mode
waves to drive field line resonances exterior to their turning points,
i.e. in the region where they are propagating. Here we find a similar
effect for magnetopause surface modes. The location in which the surface
wave frequency matches the Alfv\'{e}n mode occurs within the current
layer, as previously suggested by \citet{kozyreva19}. The turning
point of the wave, beyond which it becomes evanescent, however, occurs
outside of the boundary layer itself within the magnetosphere.
\item Realistic magnetic geometries introduce additional nodes to perpendicular
magnetic field oscillations that are not present in the velocity.
The nodes occur at the apexes of field lines, particularly near the
cusps, since large-scale radial displacements of the field line at
these locations will not cause deflection of the magnetic field vectors.
These effects are not present in box models since they contain no
such apexes as field lines are straight.
\item A reversal in the compressional magnetic field occurs at high latitudes
near the cusps due to the non-uniform magnetic field. This occurs
even when the plasma displacement has fundamental standing structure
along the field and thus nodes only at the ionospheres. We show that
both dipole and global MHD magnetic fields predict such a reversal
of the compressive oscillations due to the gradients of scale factors
and field strengths present. Only waves with short radial extents
($\ll1\text{\textendash}2\,\mathrm{R_{E}}$) are likely unaffected.
\item We report on the velocity polarizations associated with stationary
surface waves subject to a non-zero magnetosheath flow. For zero external
flow, no handedness is predicted since perturbations are a simple
breathing motion back and forth \citep{lamb32}. Stationary waves
are possible under non-zero magnetosheath flows via the surface wave
propagating against the flow, balancing its advective effect (A21).
But surface waves transfer momentum to particles in the direction
of propagation, causing them to undergo cycloidal motion \citep{stokes1847}.
This, therefore, results in a handedness to the polarization ---
right-handed with respect to the field in the pre-noon sector and
left-handed post-noon. Once advection overcomes the wave propagation
sweeping the waves tailward, outside of the $09h<\mathrm{MLT}<15h$
range, the usual sense of polarization is recovered \citep{southwood68,samson71}.
Therefore, velocity (or equivalently electric field) polarizations
measured by spacecraft may be a useful technique in detecting magnetopause
surface eigenmodes.
\item In the outer magnetosphere, close to the magnetopause, the polarization
of the magnetic field has opposite handedness to that of the velocity
due to geometric effects of the cusps. Local maxima in the radial
scale factors occur away from the magnetic equator and towards the
cusps for these field lines, unlike in a dipole field line. However,
azimuthal scale factors still decrease away from the equator. Therefore,
the gradient along the field of the scale factors is opposite for
the two directions between these local maxima. This results (in the
long wavelength limit applicable here) in the observed opposite handedness
of the magnetic field. Polarizations measured from the ground, however,
are not affected and thus they may be used in diagnosing magnetospheric
normal modes.
\item Widely-used detection techniques for standing structure both across
\citep{waters02} and along \citep{kokubun77,takahashi84} the field
can fail in a realistic magnetosphere. These make assumptions, based
on axially symmetric models, on the directions in which to compute
cross-phases between quantities. We show that in a realistic magnetosphere
they are not always the appropriate directions to use and that a method
which considers all directions perpendicular to the background magnetic
field is required.
\end{enumerate}
While we have only focused on one frequency range within this simulation,
we conclude that these effects occur when the characteristic spatial
scales of waves are much longer than those of changes in the geometry
or magnetic field. Eigenfrequencies of the MHD wave modes depend on
the Alfv\'{e}n speeds throughout the system, thus for different conditions
similarly large-scale ($>2\,\mathrm{R_{E}}$) waves will occupy different
frequency ranges \citep{archer15,archer15b,archer17}. Therefore,
our results should be applicable beyond simply the frequency range
presented.

Fully exploring the implications of these results on energy transfer
throughout geospace warrants dedicated study, though we briefly discuss
their possible impacts. It is clear from equation~\ref{eq:poynting-phasor}
that the changes to the surface mode's magnetic field perturbations
introduced at high latitudes affects the waves' energy flux. This
should be most significant along the field, likely increasing dissipation
both in the boundary layer \citep{chen74} and ionosphere \citep{allan82,southwood00}.
Additionally, our results suggest bouncing radiation belt particles
are subject to more compressional wave power, at high latitudes, than
would be expected from box models. This could lead to enhanced radial
diffusion \citep[e.g.][]{elkington06}.

The simulation presented here offers a more representative magnetosphere
than box or axially symmetric dipole models of ULF waves. However,
there are further improvements that could make the magnetosphere even
more realistic. Firstly, the run presented is perfectly North--South
and dawn--dusk symmetric due to the use of a fixed dipole with zero
tilt, no plasma corotation, and perfectly northward IMF. Secondly,
the uniform ionospheric conductivity used in unrealistic and could
be improved to include the auroral oval and inter-hemispheric differences
\citep[e.g.][]{ridley04,ridley06}. The introduction of asymmetries
to the system could thus be studied. 

No plasmasphere or ring current were included, since our focus was
the outer magnetosphere. These would lead to non-monotonic wave speed
profiles that might enable fast magnetosonic waves to penetrate the
magnetosphere more deeply \citep[e.g.][]{claudepierre16} as well
as introduce the possibility of plasmaspheric cavity modes \citep[e.g.][]{waters00},
neither of which should affect our conclusions.

Finally, we note the version of BATS-R-US uses an isotropic pressure.
This can cause unphysical mixing in collisionless space plasmas between
parallel and perpendicular pressures, affecting the magnetosonic wave
modes. While low-$\beta$ magnetospheric plasmas will be little affected,
and the results presented have all been reconciled with theory, there
may in reality be differences in high-$\beta$ areas such as the magnetosheath
or cusps. Incorporating pressure anisotropy could be investigated,
though the typical Chew-Golderberger-Low (CGL) kinetic approximation
of the MHD equations often applied to simulations \citep{chew56,meng12}
might not be appropriate for the low frequencies under consideration
since the accessible volume to particles becomes essentially the entire
flux tubes. It is conceivable that appropriately modelling the cusps
though might reveal a hitherto unforeseen eigenmode corresponding
to magnetosonic (or sound) waves trapped within the cone-like cavity
of each magnetospheric cusp, a challenge we leave to future work.

\clearpage{}

\clearpage{}

\appendix

\section{Dipole magnetic field\label{sec:dipole}}

In a dipole magnetic field, field lines for a given $L$-shell are
given by

\begin{equation}
r=L\sin^{2}\theta
\end{equation}
where $r$ is the geocentric distance and $\theta$ the colatitude.
Their length is \citep{chapman56}
\begin{equation}
s=\frac{2L}{\sqrt{3}}\left\{ \sinh^{-1}\left[\sqrt{3\left(1-\frac{1}{L}\right)}\right]+\sqrt{3\left(1-\frac{1}{L}\right)\left(4-\frac{3}{L}\right)}\right\} 
\end{equation}
and the field strength is

\begin{equation}
B_{0}=B_{E}\left(\frac{R_{E}}{r}\right)^{3}\sqrt{1+3\cos^{2}\theta}
\end{equation}
where $B_{E}=3.12\times10^{-5}\,\mathrm{T}$ is its equatorial value
on Earth's surface. One possible scheme of dipole coordinates uses
perpendicular radial and azimuthal coordinates (and corresponding
scale factors) given by \citep{swisdak06,wright20}
\begin{equation}
\begin{array}{cc}
\begin{array}{ccc}
r_{\perp} & = & \frac{r}{\sin^{2}\theta}\\
h_{r_{\perp}} & = & \frac{\sin^{3}\theta}{\sqrt{1+3\cos^{2}\theta}}
\end{array} & \begin{array}{ccc}
\phi_{\perp} & = & \phi\\
h_{\phi_{\perp}} & = & r\sin\theta
\end{array}\end{array}
\end{equation}
From these definitions, it can be shown that
\begin{equation}
\frac{\partial h_{\phi_{\perp}}}{\partial r_{\perp}}=\frac{\partial h_{\phi_{\perp}}}{\partial r}\left(\frac{\partial r_{\perp}}{\partial r}\right)^{-1}+\frac{\partial h_{\phi_{\perp}}}{\partial\theta}\left(\frac{\partial r_{\perp}}{\partial\theta}\right)^{-1}=\frac{1}{2}\sin^{3}\theta
\end{equation}
and
\begin{equation}
\frac{\partial B_{0}}{\partial r_{\perp}}=\frac{\partial B_{0}}{\partial r}\left(\frac{\partial r_{\perp}}{\partial r}\right)^{-1}+\frac{\partial B_{0}}{\partial\theta}\left(\frac{\partial r_{\perp}}{\partial\theta}\right)^{-1}=-\frac{12B_{0}}{r}\frac{\cos^{2}\theta}{1+3\cos^{2}\theta}
\end{equation}

\section{Polarization parameters\label{sec:Polarisation-parameters}}

Parameters of the polarization ellipse are determined from power spectra
\citep{arthur76,collett05} as detailed here. Firstly, the auxilliary
angle $\eta$ is calculated via
\begin{equation}
\tan\eta=\sqrt{\frac{P_{yy}}{P_{xx}}}
\end{equation}
where $P$ represents the power spectral density matrix with $x$
and $y$ being two orthogonal directions that are both perpendicular
to the background magnetic field. Combining this with the cross-phase
$\Delta$, the phase difference between the oscillations in the $y$
and $x$ directions, gives the ellipticity

\begin{equation}
\epsilon=\frac{\sin2\eta\cdot\sin\Delta}{2\sqrt{1-\nicefrac{1}{4}\sin^{2}2\eta\cdot\sin^{2}\Delta}}
\end{equation}
which has values between $-1$ (right-hand circularly polarized) and
$+1$ (left-hand circularly polarized). The orientation angle $\psi$
that the semi-major axis makes with the $x$ direction is given by
\begin{equation}
\sin2\psi=\tan2\eta\cdot\sin\Delta
\end{equation}
We also calculate the four \citet{stokes1852} parameters 
\begin{equation}
\begin{array}{ccc}
S_{0} & = & P_{xx}+P_{yy}\\
S_{1} & = & P_{xx}-P_{yy}\\
S_{2} & = & 2\mathrm{Re}\left\{ P_{xy}\right\} \\
S_{3} & = & -2\mathrm{Im}\left\{ P_{xy}\right\} 
\end{array}
\end{equation}
where $S_{0}$ is the total variance, $S_{1}$ and $S_{2}$ correspond
to linearly polarized waves, and $S_{3}$ to circularly polarized
waves \citep{collett05}. The degree of polarization $p$ is thus
\begin{equation}
p=\frac{\sqrt{S_{1}^{2}+S_{2}^{2}+S_{3}^{2}}}{S_{0}}
\end{equation}
which varies between $0$ (unpolarized) and $1$ (fully polarized).

\begin{notation}

\notation{$msh$}Magnetosheath

\notation{$mp$}Magnetopause

\notation{$sph$}Magnetosphere

\notation{$GSM$}Geocentric Solar Magnetospheric coordinates

\notation{$\parallel$}Field-aligned

\notation{$\alpha$}Arbitrary perpendicular coordinate

\notation{$\beta$}Arbitrary perpendicular coordinate

\notation{$\epsilon$}Ellipticity

\notation{$\theta$}Colatitude

\notation{$\phi$}Azimuthal angle

\notation{$\phi_{\perp}$}Perpendicular azimuthal coordinate

\notation{$\mu_{0}$}Vacuum permeability

\notation{$\xi$}Plasma displacement

\notation{$\rho$}Mass density

\notation{$\omega$}Angular frequency

\notation{$\mathbf{B}$}Magnetic field

\notation{$c_{s}$}Speed of sound

\notation{$\mathbf{e}$}Orthonormal basis vector

\notation{$\mathbf{E}$}Electric field

\notation{$h$}Curvilinear scale factor

\notation{$\mathbf{J}$}Current density

\notation{$\mathbf{k}$}Wave vector

\notation{$r_{\perp}$}Perpendicular radial coordinate

\notation{$\mathbf{r}$}Geocentric Position

\notation{$s$}Field line length

\notation{$\mathbf{S}$}Poynting vector

\notation{$t$}Time

\notation{$\mathbf{v}$}Plasma velocity

\notation{$v_{A}$}Alfv\'{e}n speed

\end{notation}

\section*{Open Research}
Simulation results have been provided by the Community
Coordinated Modeling Center (CCMC) at Goddard Space Flight Center
using the SWMF and BATS-R-US tools developed at the University of
Michigan's Center for Space Environment Modeling (CSEM). This data
is available at \url{https://ccmc.gsfc.nasa.gov/results/viewrun.php?domain=GM&runnumber=Michael_Hartinger_061418_1}.

\begin{acknowledgments}
The authors thank Tom Elsden for helpful discussions regarding this
work. MOA holds a UKRI (STFC / EPSRC) Stephen Hawking Fellowship EP/T01735X/1.
DJS was supported by STFC grant ST/S000364/1. MDH was supported by
NASA grant 80NSSC19K0127. A.N.W. was partially funded by STFC grant
ST/N000609/1.
\end{acknowledgments}

\end{document}